\documentclass[epj]{webofc}
\usepackage[varg]{txfonts}   
\usepackage{subfigure}
\usepackage{color}
\woctitle{MESON2016 - the 14$^\textrm{th}$ International Workshop on Meson Production, Properties and Interaction}
\begin{document}
\selectlanguage{english}
\title{Amplitude analysis of resonant production in three pions}

\author{Andrew~Jackura\inst{1}\fnsep\thanks{\email{ajackura@indiana.edu},} \and
        Mikhail~Mikhasenko\inst{2} \and
        Adam~Szczepaniak\inst{1,3} 
       \\for the JPAC Collaboration
}

\institute{Indiana University, Bloomington IN, U.S.A.
\and
           Universit\"at Bonn, Helmholtz-Institut f\"ur Strahlen- und Kernphysik, Bonn, Germany
\and
           Thomas Jefferson National Accelerator Facility, Newport News VA, U.S.A.
          }

\abstract{
  We present some results on the analysis of three pion resonances. The analyses are motivated by the recent release of the largest data set on diffractively produced three pions by the COMPASS collaboration. We construct reaction amplitudes that satisfy fundamental $S$-matrix principles, which allows the use of models that have physical constraints to be used in fitting data. The models are motivated by the isobar model that satisfy unitarity constraints. The model consist of a Deck production amplitude with which final state interactions are constrained by unitarity. We employ the isobar model where two of the pions form a quasi-stable particle. The analysis is performed in the high-energy, single Regge limit. We specifically discuss the examples of the three pion $J^{PC}=2^{-+}$ resonance in the $\rho\pi$ and $f_2\pi$ channels.
}
\maketitle
\section{Introduction}
\label{intro}
We aim to study the peripheral resonance production of $3\pi$ systems. Motivated by the large data set at COMPASS, we construct a reaction amplitude model for $\pi^- p\to \pi^-\pi^-\pi^+p$ at high energies. The model will be used in a mass-dependent analysis of the COMPASS partial wave intensities in \cite{Adolph:2015tqa}. The COMPASS results provide high statistics, high purity data which allows for a detailed analysis. The Joint Physics Analysis Center (JPAC) is affiliated with COMPASS to perform these analysis.

The model is constructed to satisfy fundamental principles of $S$-matrix theory. We impose unitarity on the reaction amplitudes, and ensure that the model obeys the proper analytic constraints required by causality. We emphasize the production process as an important role in determining physics. The $\pi^-\pi^-\pi^+\equiv(3\pi)^-$ system is produced by a particle exchange in between the incident pion and proton, as well as an exchange between the incident pion and the reggeon that is exchanged in the full reaction. This production process is known as the Deck production process, and has been studied as possible explanations for various resonances \cite{Ascoli:1974sp, Basdevant:2015wma, Basdevant:1977ya}.

Peripheral production is advantageous as it allows for the factorization of the amplitude into two parts, one being the effective $2\to n$ meson-Reggeon scattering amplitude, where $n$ is the number of produced mesons. We develop the model for any $J^{PC}$ quantum number, but focus first on the $J^{PC}=2^{-+}$ sector. The $2^{-+}$ sector gives us the opportunity to study the long standing puzzle of the $\pi_2(1670)-\pi_2(1880)$ interplay. These mesons have the exact same quantum numbers, but are only $\sim 200$ MeV apart, which disagrees with Regge trajectories of the mesons.

\section{Formalism}
\label{sec-1}

The reaction model exhibits numerous features. We want to model the $\pi^-p\to(3\pi)^-p$ reaction in the high energy regime to compare to COMPASS. At high energies, the reaction is dominated by pomeron exchange. At high energies, we can use factorization and effectively separate the nucleon-pomeron-nucleon vertex from the the $\pi^-\mathbb{P}\to\pi^-\pi^-\pi^+$ amplitude. We work in the quasi-two body limit, where the $\pi^-\pi^+$ mesons form a quasi-stable neutral isobar of spin $S$ denoted $I_S^0$. This leads to the factorization $\pi^-\mathbb{P}\to I_S^0\pi^-$ and $I_S^0\to\pi^-\pi^+$. We model the pomeron phenomenologically as having a vector coupling since the Regge pole for the pomeron is $\alpha_{\mathbb{P}}\approx 1$. The amplitude for the total reaction is denoted by $A(s,\Omega,\Omega')$, where $s$ is the invariant mass-squared of the $(3\pi)^-$ system, $\Omega$ and $\Omega'$ are the orientations of the isobar and one of the pions in the isobar respectively, and all other kinematic variables are suppressed for notational convenience. The partial wave decomposition is 
\begin{equation}\label{eq:ampPWdecomp}
A(s,\Omega,\Omega') \sim \sum_{i}F_i(s)\,Z_i(\Omega,\Omega'),
\end{equation}
where $F_i(s)$ are the partial wave amplitudes for the $i^{\text{th}}$ channel, and $Z(\Omega,\Omega')$ contains all the rotational dependence of the system for each channel. The channel index $i$ indicates the channel with $(3\pi)^-$ quantum numbers $I^{G}(J^{PC})M^{\epsilon}\,\,I^0_S\pi\,\,L$, where $I$ is the total isospin, $G$ is the $G$-parity, $J$ is the total angular momentum, $P$ is the parity, $C$ is the $C$-parity, $M$ is the total spin projection, $\epsilon$ is the reflectivity, and $L$ is the angular momentum between the isobar $I_S^0$ and the spectator pion. For the $(3\pi)^-$ system, $I=1$, $G=-$, and $C=+$ always, and for pomeron exchange $\epsilon=+$.

\subsection{Unitarity and analyticity}
\label{sec-2}

Unitarity of the $S$-matrix puts constraints on the form of amplitudes. We impose these constraints by requiring the partial wave amplitudes satisfy
\begin{align}
\Delta\,F_{i}(s) = 2i\,\sum_{j}t^{*}_{ij}(s)\,\rho_{j}(s)\,F_{J}(s) \label{eq:unitF}\\
\Delta\,t_{ij}(s) = 2i\,\sum_{k}t^{*}_{ik}(s)\,\rho_k(s)\,t_{kj}(s) \label{eq:unitT},
\end{align}
where $t_{ij}(s)$ is the rescattering amplitude of the isobar and spectator pion, $\rho_j(s)$ is the quasi-two body phase space, and $\Delta$ indicates the discontinuity of the amplitude across the unitarity cut. It is assumed the $F_i(s)$ has two cuts, the unitarity cut from the $3\pi$ threshold to $\infty$, and a left hand cut generated by the production process. We can write a dispersive integral for Eq~\eqref{eq:unitF} which has a solution
\begin{equation}\label{eq:Famp}
F_{i}(s) = b_{i}(s) + \sum_{j}t_{ij}(s)\,c_{j}+\sum_{j}t_{i}(s)\int_{s_{th}}^{\infty}ds'\,\frac{\rho_{j}(s')\,b_{j}(s')}{s'-s}.
\end{equation}
Here, $b_{i}(s)$ is the projection of the Deck production amplitude, $c_{j}$ is a vector of couplings related to the short range production of the three pions, and the last term is needed to satisfy unitarity, often called the unitarized Deck amplitude. Note that in our amplitude, $b_{i}(s)$ is a model, but parameters $c_{j}$ will be fit parameters.

We must parameterize the rescattering amplitude in such a way to satisfy Eq.~\eqref{eq:unitT}. We choose to use the $K$-matrix parameterization
\begin{equation}\label{eq:Kmat}
[t^{-1}]_{ij}(s) = [K^{-1}]_{ij}(s) + I_{i}(s)\delta_{ij},
\end{equation}
where $K_{ij}(s)$ is a real-symmetric matrix, and $I_i(s)$ is the Chew-Mandelstam phase space factor satisfying $\text{Im}\,I_i(s) = -\rho_i(s)$. In this model, it is assumed that $t_{ij}(s)$ has only the unitarity cut. It is also assumed that all resonance content comes from this amplitude. therefore, we must parameterize the $K$-matrix such that we generate poles on the unphysical Riemann sheets. We try various parameterizations of the $K$-matrix, and look for the corresponding resonance in systematic studies. The fit parameters are the parameters in the $K$-matrix, as well as the production vector $c_j$.

\subsection{Production mechanism}
\label{sec-3}
The closest left hand cut to the physical region comes from $\pi$-exchange in the $\pi^-\mathbb{P}\to I_{S}^0\pi^-$ amplitude. The production process is modeled by the amplitude
\begin{equation}\label{eq:DeckAmp}
A_{Deck}(s,\Omega) = \frac{g_{I\pi\pi}g_{\mathbb{P}\pi\pi}}{t(s,\theta)-m_{\pi}^2}\epsilon_{\lambda}\cdot p_2\epsilon^{S*}_{\lambda'}\cdot \{p_a\},
\end{equation}
where $g_{I\pi\pi}$ and $g_{\mathbb{P}\pi\pi}$ are couplings for $\pi\pi$ to the isobar and pomeron respectively, $t(s\theta)$ is the invariant momentum transfer between the incident pion and the pomeron, and $\epsilon_\lambda$ and $\epsilon_{\lambda'}^{S}$ are the polarization tensors of the pomeron and isobar respectively. The four-momenta $p_a$ and $p_2$ are of the incident and spectator pion respectively. Note the symbol $\{p_a\}$ indicates that there will be $S$ contractions of the momentum to the polarization tensor. The couplings are taken from known data on various reactions involving these processes. Therefore, the Deck amplitude is completely determined, and is used as an input for our model. The partial wave projection of the Deck amplitude $b_i(s)$ is given by projecting out $A_{Deck}$ with the appropriate rotational dependencies.

\section{Current progress and future work}
At the current state, the formalism has been completed and systematic tests and preliminary fits are underway. It is proposed that the model is first fitted to the partial wave intensities of the $2^{-+}$ sector from the data gathered by the COMPASS collaboration. Once final fits are obtained, the resonance content will be extracted from the analytically continued amplitudes. We construct intensity distributions 
\begin{equation}\label{eq:intensity}
I_i(s) = \rho_i(s) \lvert F_{i}(s)\rvert^2,
\end{equation}
which are fit to the partial wave intensity distributions found in \cite{Adolph:2015tqa}. We also fit relative phase information between different channels.

\subsection{Resonance pole extraction}
\begin{figure}[ht]
\centering
\begin{subfigure}[\ Two-body phase space]{
\includegraphics[width=0.4\columnwidth]{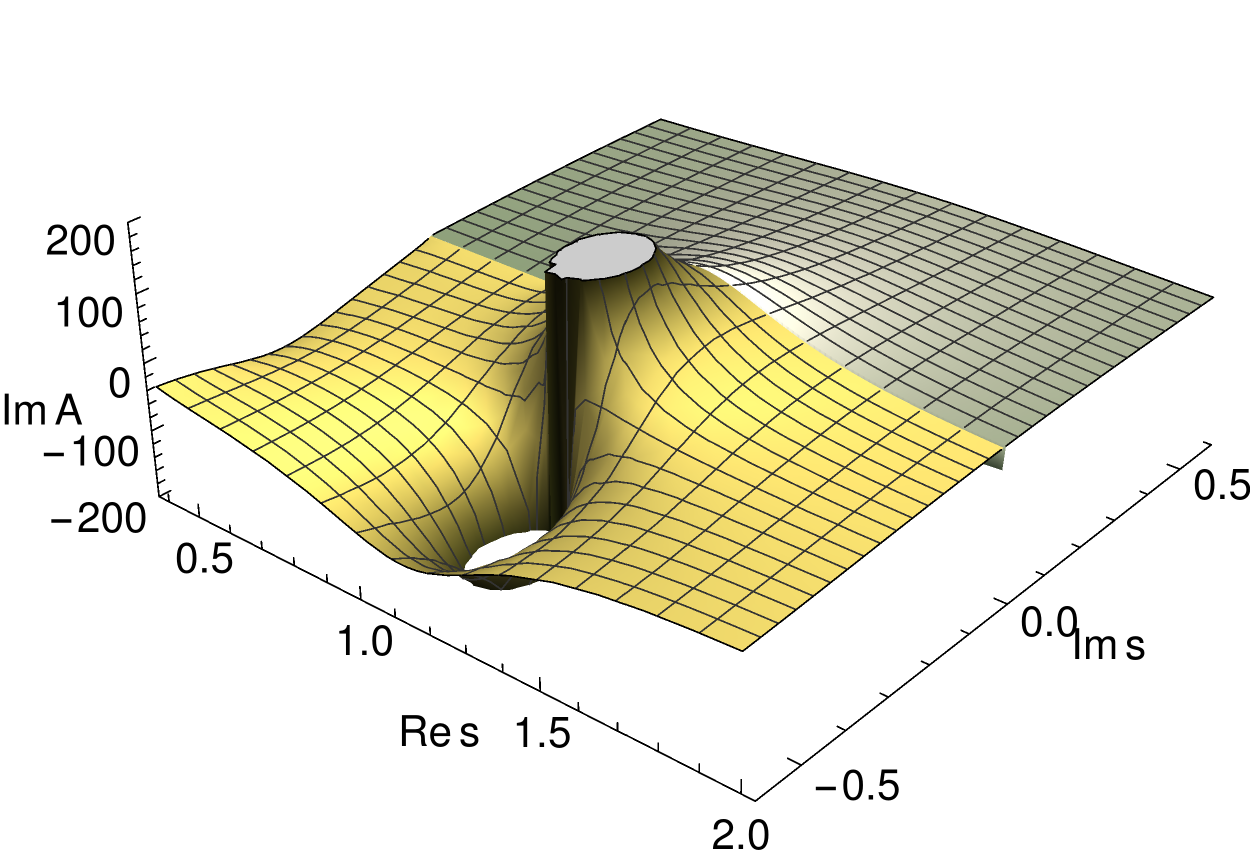}}
\label{fig:2body}
\end{subfigure}
\begin{subfigure}[\ Quasi two-body phase space]{
\includegraphics[width=0.4\columnwidth]{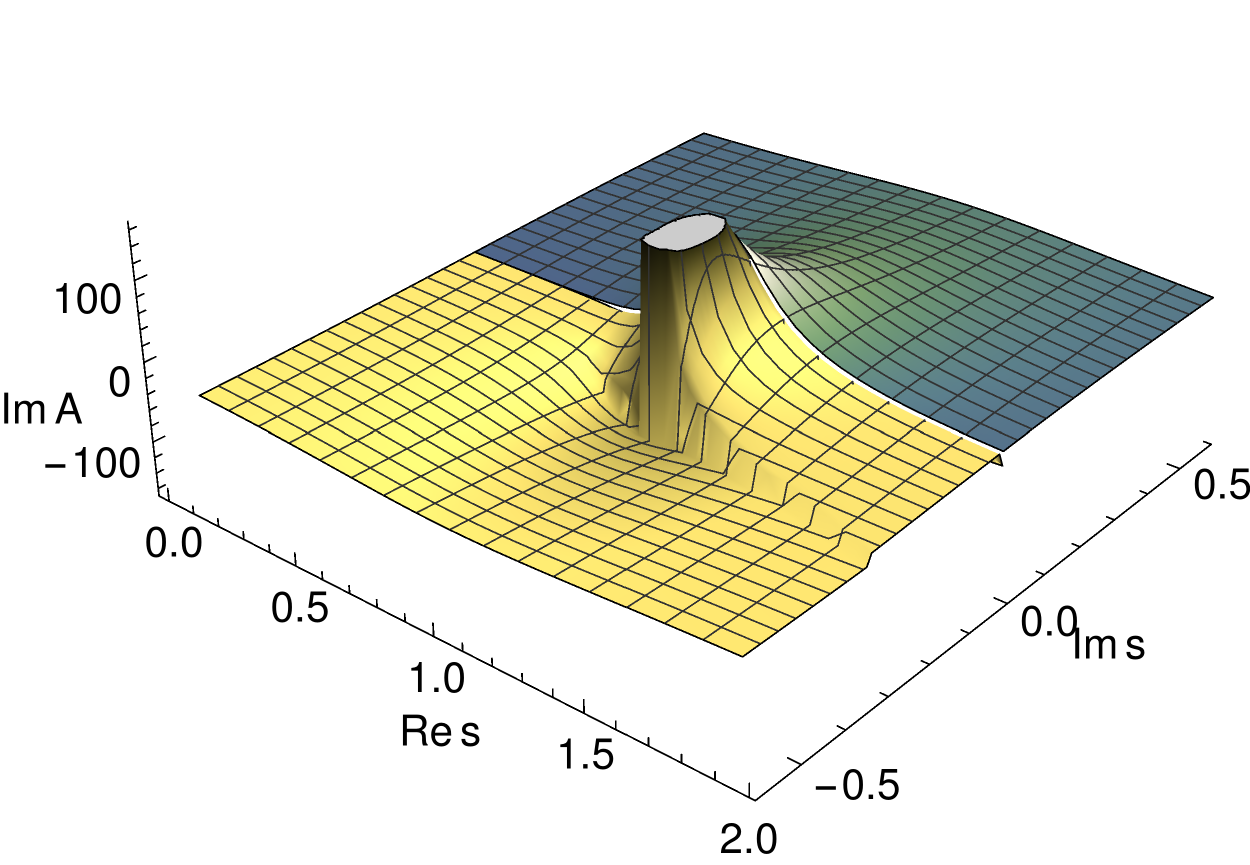}}
\label{fig:3body}
\end{subfigure}
\caption{Complex $s$-plane of $\text{Im}\,A$, where $A$ is an amplitude with some resonance. The green section is the first sheet while the yellow section is the second sheet. In (a), we show what the typical pole structure is in the case of a simple two-body phase space factor. In (b), additional cuts appear due to the unstable isobar. Notice that the resonance pole is hidden underneath the cut.} 
\label{fig:PoleSearch}
\end{figure}

Resonance content from amplitudes is determined by looking for poles of $t_{ij}(s)$ on the unphysical Riemann sheets. Once the parameters are fitted to the COMPASS partial waves, the rescattering amplitude is analytically continued to the second sheet and we look for resonances. The amplitude on the second sheet is $[t_{\text{II}}^{-1}]_{ij}(s) = [t^{-1}]_{ij}(s) - 2i\rho_i(s)\delta_{ij}$. We need to discuss some complications due to the quasi two-body phase space factor. The quasi two-body phase space factor contains the isobar information, which leads to cuts on the unphysical sheet known as Woolly cuts. Resonance may appear under these cuts, which requires the knowledge of how to continue the amplitude below these cuts. Fig.~\ref{fig:PoleSearch} illustrates this with a simple resonance example. The typical two-body phase space factor is given by $\rho\sim 2k/\sqrt{s}$, where $k$ is the relative momentum between the isobar and the spectator pion. If the isobar decays, the phase space is modified to the quasi two-body phase space $\rho_Q$,
\begin{equation}
\rho_Q\sim\int ds' \rho(s')\,\text{Im}\,f(s'),
\end{equation}
where $\rho$ is the typical two-body phase space, and $f(s)$ is the isobar shape.

\section{Summary}
We have developed the formalism to analyze $3\pi$ system for peripheral reactions. Our model satisfies the unitarity and analyticity requirements of $S$-matrix principles, allowing for more rigorous extraction of resonances. We are applying the formalism to the COMPASS results on $\pi^-p\to(3\pi)^-p$ in the $2^{-+}$ sector. As of the writing of this document, the formalism has been completed, and systematic fits are ongoing. Once fits are complete, we will search for resonances of the $(3\pi)^-$ system. In future studies, we can apply the model to other $J^{PC}$ sectors, and it is of interest to extend the model for photon beams to be used for the JLab12 upgrade.

\begin{acknowledgement}
This work was supported in part by  U.S. Department of Energy, Office of Science, Office of Nuclear Physics under contract DE-AC05-06OR23177 and by the U.S. Department of Energy under Grant No. DE-FG0287ER40365.
\end{acknowledgement}
%
%
%

\end{document}